\expandafter\ifx\csname natexlab\endcsname\relax\fi
\providecommand{\url}[1]{\href{#1}{#1}}
\providecommand{\dodoi}[1]{doi:~\href{http://doi.org/#1}{\nolinkurl{#1}}}
\providecommand{\doeprint}[1]{\href{http://ascl.net/#1}{\nolinkurl{http://ascl.net/#1}}}
\providecommand{\doarXiv}[1]{\href{https://arxiv.org/abs/#1}{\nolinkurl{https://arxiv.org/abs/#1}}}

\def \aap  {A\&A}

\def \apj  {ApJ}

\def \apjl  {ApJL}

\def \mnras {MNRAS}

\def \pasa {PASA}
\def \prl {PRL}

\documentclass[a4paper,11pt]{article}
\usepackage{pos}

\title{Maximum energy achievable in supernova remnants: self-consistent simulations}
 \ShortTitle{The Maximum Energy CRs from SNRs}

\author*[a]{Emily Simon}
\author[a,b]{Damiano Caprioli}
\author[c]{Colby Haggerty}
\author[d]{Brian Reville}

\affiliation[a]{The University of Chicago, Department of Astronomy and Astrophysics,\\
5640 S Ellis Ave, Chicago, IL 60637, USA}

\affiliation[b]{The University of Chicago, Enrico Fermi Institute,\\
5640 S Ellis Ave, Chicago, IL 60637, USA}

\affiliation[c]{University of Hawaii, Institute for Astronomy,\\
Honolulu, HI 96822, USA}

\affiliation[d]{Max-Planck-Institut f¨ur Kernphysik,\\
Saupfercheckweg 1, Heidelberg, 69117, Germany}

\emailAdd{ersimon@uchicago.edu}
\emailAdd{caprioli@uchicago.edu}
\emailAdd{colbyh@hawaii.edu}
\emailAdd{brian.reville@mpi-hd.mpg.de}

\abstract{It has been long believed that oblique and quasi-perpendicular configurations in supernova remnants (SNRs) were inefficient at injecting ions into diffusive shock acceleration (DSA), and that the highest energy Galactic cosmic rays (CRs) must come from parallel or quasi-parallel shocks. 
However, recent 3D kinetic simulations have shown that high-obliquity shocks can successfully energize particles and produce amplified magnetic fields in the upstream. We aim to investigate the maximum energy particles it is possible to produce in oblique and quasi-perpendicular shocks and whether they are capable of triggering the non-resonant hybrid instability (NRHI). 
We present a novel setup for hybrid simulations of non-relativistic shocks that use a ``faux shock'' boundary condition instead of a real shock to significantly reduce the computational cost and that can be run for long enough to study the late-time behaviors of these systems. 
Our results show that it may be possible for oblique and quasi-perpendicular shocks to transition from early periods of shock drift acceleration (SDA) into DSA at later times, giving particles a brief period of rapid acceleration followed by a long-duration, self-sustaining period of slower energy growth. Furthermore, we find evidence that the NRHI is triggered in the upstream at late times. Oblique and quasi-perpendicular shocks may be an important contributor to high energy CRs, potentially even responsible for the knee in the CR energy spectrum.}

\ConferenceLogo{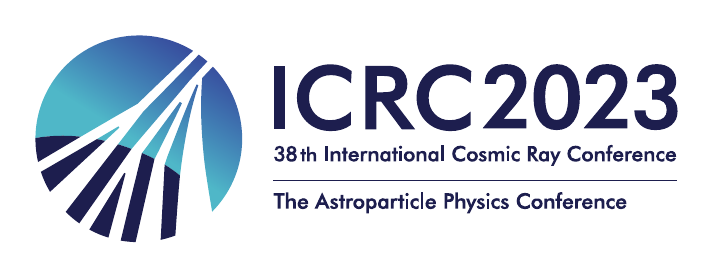}

\FullConference{%
38th International Cosmic Ray Conference (ICRC2023)\\
  26 July - 3 August, 2023\\
  Nagoya, Japan}

\begin{document}
\maketitle

\section{Introduction}

There is good reason to believe that cosmic rays (CRs) with energies $E \lesssim 10^{15} eV$ have galactic origins (\cite{Hillas2005}; \cite{Berezhko2007}; \cite{Ptuskin2010}; \cite{Caprioli2010}). Of the known candidate sources within the galaxy, supernova remnants (SNR), which accelerate particles via diffusive shock acceleration (DSA), are expected to produce CRs up to energies of a few PeV (\cite{Blummer2009}). 
However, analytical estimates of the maximum energy produced in DSA at shocks (\cite{Bell2013}) struggle to account for the energy needed to explain the spectral break around $4$ PeV called the ``knee". 
Furthermore, observations of known SNRs, which provide indirect evidence of CR energies from the gamma rays produced in hadronic collisions, infer maximum proton energies $\sim 100$ TeV, roughly a factor of ten too small to form the knee (\cite{Cristofari2021}). 
This implies that either galactic SNRs may not be the dominant sources of PeV CRs, or that there is something about the acceleration process that is still not well understood.

In this paper, we examine particle acceleration at oblique and quasi-perpendicular shocks with a novel and computationally efficient simulation method. 
Recent 3D kinetic simulations \cite{Orusa2023} have found that the believed failure of quasi-perpendicular shocks to re-inject particles from the downstream is an artifact of 2D simulations, in which compressed magnetic field lines in the downstream are artificially stacked and their vorticities suppressed, preventing magnetic field line wandering (\cite{Rechester1978}).
Full 3D treatment shows that weakly-magnetized oblique and quasi-perpendicular shocks do in fact re-inject particles at efficiencies which scale with Mach number and which make these configurations relevant to the story of high energy particle production. 

The dominant acceleration mechanism of oblique and quasi-perpendicular shocks early on is shock drift acceleration (SDA) which causes rapid particle acceleration $E \propto t^2$ (\cite{Chen1975}; \cite{Ball2001}). 
We study whether SDA may eventually give way to DSA at late times, meaning an initially highly-energized population of particles is created via SDA and injected into self-sustaining, long-duration DSA, allowing for potentially high energy and high number density CR production.

Additionally, in the regime where DSA can occur, we investigate the possibility of triggering the non-resonant hybrid (aka Bell) instability \cite{Bell2004,Bell2013} in which CR currents cause nonlinear magnetic field amplification conducive to a larger CR maximum energy.

\begin{figure*}
    \centering
    \includegraphics[width=1\linewidth]{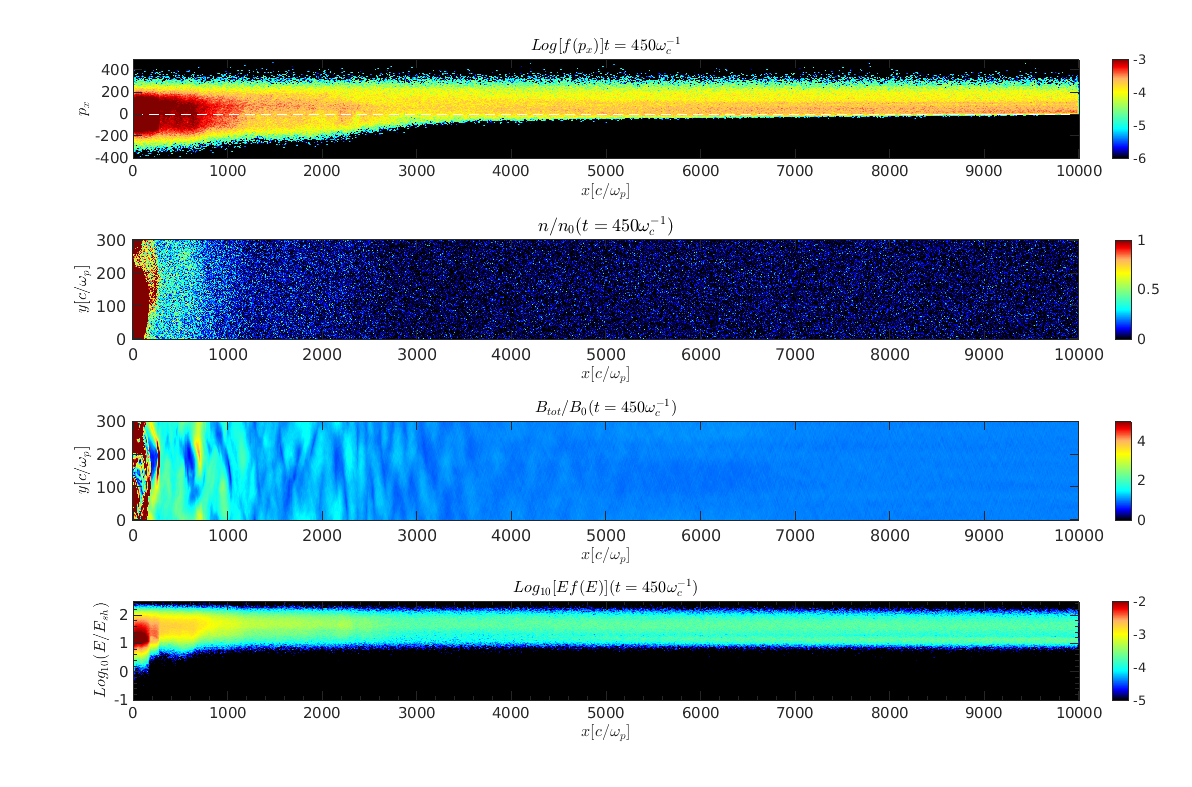}
    \caption{Top panel: $x-p_x$ ion phase space; second panel: CR number density, normalized to the injected density; third panel: total magnetic field normalized to the initial magnetic field; bottom panel: CR energy phase space, with energy  normalized to $E_{\rm sh}=\frac{1}{2}m v_{\rm sh}^2$.}
    \label{fig:benchmark}
\end{figure*}

\section{Novel Simulations with a Shock Boundary Condition}
We use the hybrid (fluid electrons--kinetic ions) code {\tt dHybridR} \cite{Haggerty2019} with a novel type of simulation which bypasses the need of creating a shock by instead having a shock-like boundary condition (or ``faux shock") on one edge of the simulation box. 
This faux-shock boundary condition behaves like a shock in that it reflects and transmits fixed percentages of incoming particles.
Ions returned by the faux shock into the upstream are given an isotropic distribution in the plasma frame and an energy kick consistent with a downstream scattering in the shock frame (which in the shock frame translates into an energy loss). 
Returning particles scatter in the upstream self-generated magnetic turbulence, gaining an average $\Delta E/E \sim u_1 -u_2$ per scattering, making the simulation functionally identical to a real shock.
According to DSA, the combined energy gain and probability of return from downstream lead to a $p^{(-3r/(r-1))}$ power-law spectrum in momentum \cite{Bell1978}, where $r=4$ is the shock compression ratio, a fixed value in our simulations.

The new simulation setup has three main advantages: 
1) we can inject an initial population of high-energy CRs instead of waiting for them to undergo many cycles of DSA, which allows us to study the late-time regimes of DSA during the ejecta-dominated stage (shock velocity remains constant) which are most relevant for understanding the maximum energy CRs it is possible to produce; 
2) we can effectively mimic the particle return from the downstream in oblique or quasi-perpendicular shocks using only 2D simulations instead of the previously required 3D; and
3) CRs are a different species sampled with 4 macroparticles per cell, which greatly increases their statistics.

\section{CR acceleration at oblique shocks}
We consider one benchmark simulation (run $\mathcal{B}$) against which we compare a simulations with varied angles between the magnetic field and the shock normal, referred to as the shock obliquity, $\mathcal{\theta}$. 
In run $\mathcal{B}$, we use a box with height $300 d_i$ and length $10^4 d_i$ where $d_i=c/\omega_p$ is the ion skin depth, where $c$ is the speed of light and $\omega_p = \sqrt{4\pi n e^2/m}$ is the ion plasma frequency with $m$, the ion mass, $e$ the ion charge, and $n$ the ion number density. 
The initial magnetic field is oriented normal to the shock surface ($\mathcal{\theta} = 0^{\circ}$) and parallel with the fluid velocity.
We fix the shock Alfv\'enic Mach number as $M_A=26.67$, where the Alfv\'en speed is $v_A = B/\sqrt{4\pi m n_g}$ and $c=1000 v_A$. 
Time is measured in inverse cyclotron times, $\omega_c^{-1} = mc/eB_0$, where $B_0$ is the initial magnetic field strength.

CRs are constantly injected at the faux shock with number density $n_{cr} = 0.1 n_g$ where $n_g$ is the number density of the background gas, with an initially isotropic momentum of $p_{iso} = 100 mv_A = 0.1c$. The background gas is injected from the right side of the box with velocity $v=v_{sh} = 26.67$. The simulation runs for $450 \omega_{c}^{-1}$ in order to approach magnetic field growth saturation via the NRHI.

\begin{figure}
    \centering
    \includegraphics[width=0.8\linewidth]{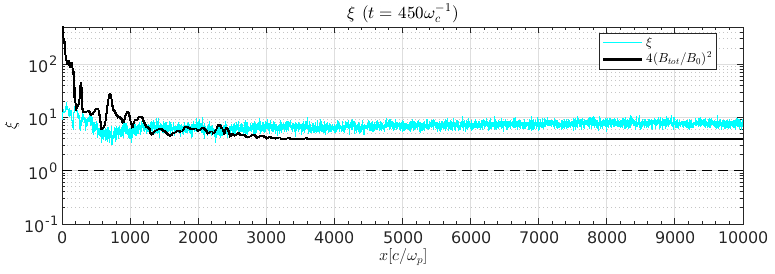}
    \caption{Local value of $\xi$ (cyan curve), the parameter that characterizes the relevance of the NRHI, for a parallel shock at $t = 450 \omega_c^{-1}$ and magnetic field amplification (black) normalized for comparison to $\xi$. At all positions  $\xi\gtrsim 1$, meaning there is sufficient CR free energy to trigger the NRHI, which should saturate at a level $(\delta B/B_0)^2\approx \xi/4$.}
    \label{fig:para_xi}
\end{figure}

The last time step from run $\mathcal{B}$ is shown in Figure \ref{fig:benchmark} which we use as a demonstration that the new simulation strategy produces the expected results for parallel shocks. 
The top panel shows the $x$-directed momentum phase space, in which the presence of particles with negative momenta clearly shows that particles are diffusing back after scattering off of self-generated magnetic fields in the upstream. The second panel of Figure \ref{fig:benchmark} shows the CR distribution spatially within the simulation box. This density gradient closely traces the magnetic field strength,  as seen in the third panel of Figure \ref{fig:benchmark}. 
The bottom panel of Figure \ref{fig:benchmark} shows the CR energy normalized to the shock energy. Here we see a high density of particles with energy $\lesssim E_{sh}$ that are confined close to the shock, whereas particles with higher energies are streaming away, consistent with the DSA picture.

Beyond this, we note that the strongly amplified magnetic fields in the upstream are consistent with the expectations from the NRHI.
Bell's modes are rapidly growing when the free energy in CRs is greater than the energy in magnetic fields \cite{Bell2004},
\begin{equation}
    \xi \approx \frac{U_{cr}v_d}{U_B c} = \rho_{cr}(u_1 +\overline{v})^2 \gtrsim 1
\end{equation}
where $u_1$ is the velocity of the upstream gas, $\overline{v}$ is the average velocity of CRs in the upstream frame, and $\rho_{cr}$ is the mass density of CRs. 
Hybrid simulations \cite{Georgios2022} show that the amplified magnetic fields should be related to the net CR energy  by 
\begin{equation}
   \left(\frac{B_{tot}}{B_0}\right)^2 \approx \frac{\xi}{4}
\end{equation}
where $B_{tot}$ is the total magnetic field strength and $B_0$ is the initial magnetic field strength. Figure \ref{fig:para_xi} shows that the value of $\xi$ throughout the simulation box is larger than $1$ and the NRHI is indeed able to grow. 
We also see a reasonable match between $\frac{\delta B}{B}$ and the predictions from $\xi$ given the time offset between a parcel of fluid's exposure to high currents and the subsequent multiple e-foldings of growth of the magnetic field in that parcel. 
In general one needs the growth time of the fastest growing mode, with growth rate $\gamma_{max} = 0.5 j_{cr}\left(4\pi/c^2 m n_g \right)$,  to be smaller than the advection time for a magnetic field fluctuation to undergo multiple e-foldings before being advected by the shock \cite{Bell2013}.

These simulations require a fixed compression ratio, and hence an injection spectrum, and we therefore rely on the findings of 3D simulations \cite{Orusa2023} to give a frame of reference to the values we have chosen for our oblique and quasi-perpendicular simulations. 
Simulation $\mathcal{O}$1 is an oblique simulation with $\theta = 60^{\circ}$, and $\mathcal{O}$2 is a quasi-perpendicular simulation with $\theta = 80^{\circ}$. 
All other simulation parameters remain the same as the benchmark case $\mathcal{B}$. We emphasize that this includes the compression ratio $r$ of all three simulations, which is always the same and set to produce the typical DSA injection spectrum of $p^{-4}$. 
While this may be an overestimate for oblique and quasi-perpendicular shocks (unless Mach number $\gtrsim$ 100, \cite{Orusa2023}), it is still useful for setting upper limits on the maximum particle acceleration we can achieve. 
We also note that as the upstream magnetic field becomes increasingly turbulent and DSA kicks in, the $p^{-4}$ spectrum becomes an appropriate choice.

\begin{figure*} [t!]
  \centering
  \includegraphics[width=0.8\linewidth]{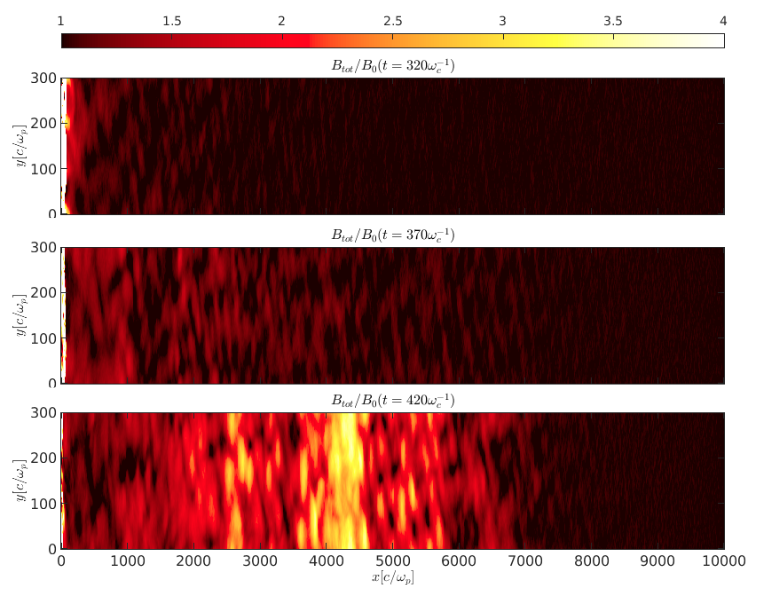}
\caption{Time evolution of the magnetic field strength for a quasi-perpendicular shock, $\theta = 80^{\circ}$.}
\label{fig:80_deg_B}
\end{figure*}

\section{Maximum energy CRs}

\begin{figure}
    \centering
    \includegraphics[width=0.7\linewidth]{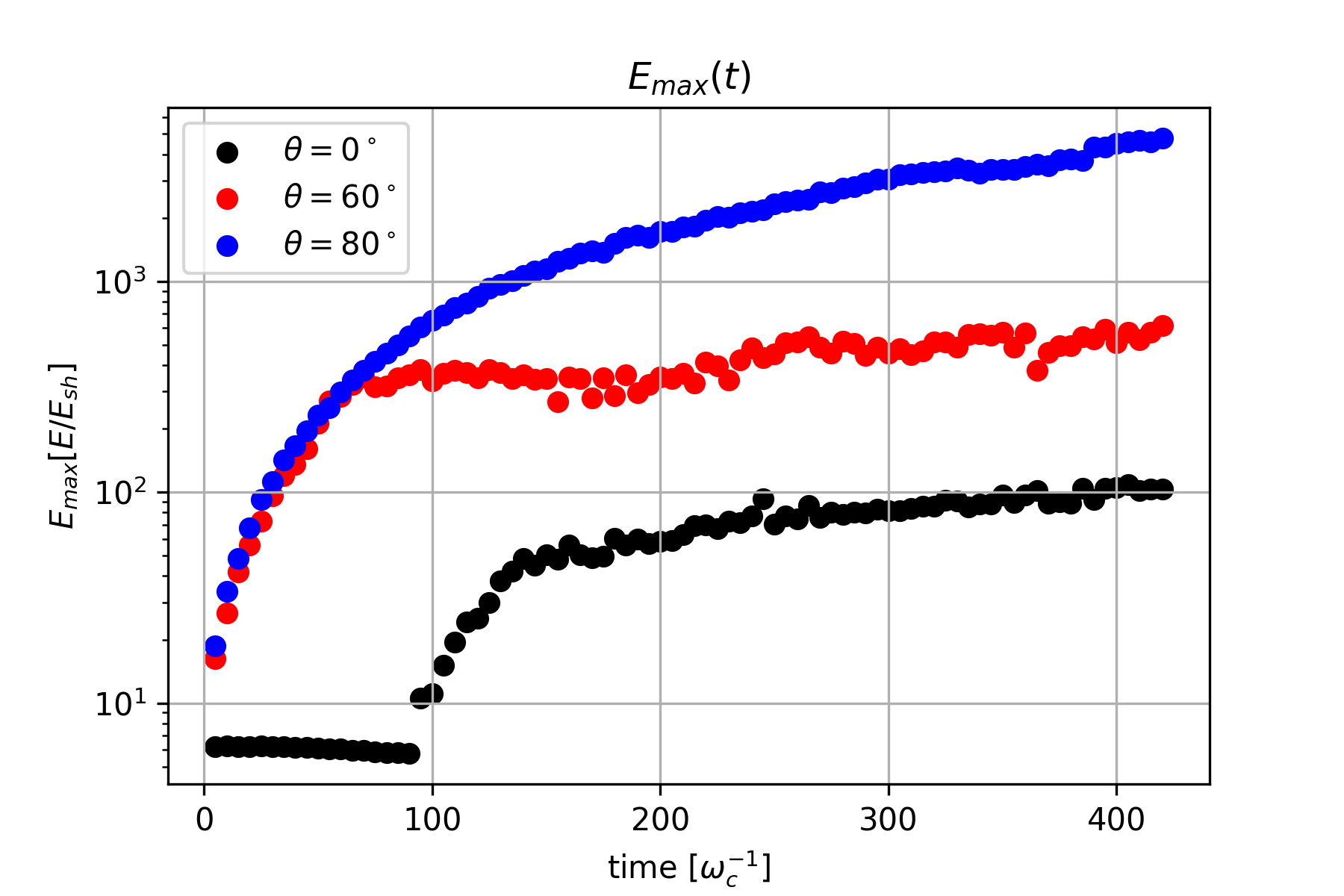}
    \caption{The maximum energy CRs within the simulation as a function of time for the parallel (black), oblique (red), and quasi-perpendicular (blue) configurations. All energies are normalized to the shock energy.}
    \label{fig:Emax_t}
\end{figure}

The highest obliquity simulation tested in this work is $\mathcal{O}2$, a quasi-perpendicular shock with $\theta = 80^{\circ}$. We find in \ref{fig:Emax_t} that the maximum energy as a function of time $E_{max}(t)$ scales as $t^2$ which is characteristic of SDA, and expected for quasi-perpendicular shocks. The scale of the energy growth reaches $5000E_{sh}$ which we note is unphysical due to our unrealistic re-injection prescription. In realistic SDA, particles spend only a short period of time undergoing acceleration before they are advected into the downstream ($\Delta t \propto \omega_c^{-1}p$), preventing such extreme energy gains (\cite{Orusa2023}). 

$\mathcal{O}2$ shows evidence that particles do escape into the upstream and the NRHI does in fact grow, however it takes significantly longer for the field to grow compared to the parallel case. This is because it takes particles longer to propagate into the $+x$ direction when they are confined to their nearly-vertical field lines. This has the effect of limiting the space, and therefore timescale, over which a magnetic perturbation started in the upstream is able to grow before being advected into the downstream. If we wait long enough, however, current from escaping CRs reaches some $x_{crit}$ wherein the current is large enough to make significant contributions to the amplified field, and the amplification started at this position has a long enough time to grow. Calculating the precise location of this position is nontrivial because it is nonlinear in time and space, but it can be estimated by checking whether a given position in a simulation box is over the threshold for the NRHI to occur, ie: if $\xi \gtrsim 1$.

In Figure \ref{fig:80_deg_B}, the top panel shows the first developments of small magnetic field amplifications in the upstream which are being advected towards the shock. The middle panel shows a later time step at $t = 370 \omega_c^{-1}$ in which the strength of field amplification is increasing. In the final panel, $t = 420 \omega_c^{-1}$, very strong, non-linear magnetic field growth is visible. The position of this strongly amplified region near the center of the simulation box and the nonlinear growth in time are strong indications that the NRHI is responsible for their production. 

The oblique simulation in this work has $\theta = 60^\circ$ and all other parameters the same as $\mathcal{B}$. Similar to the quasi-perpendicular case, it takes a longer time for particles to propagate large distances within the box because their trajectories are aligned with nearly-vertical field lines. After this time, we see continual evidence of the NRHI, which remains above the saturation limit for the duration of the simulation. 

The amplitude of the magnetic field strength in this oblique case is smaller than that found in the quasi-perpendicular case, despite the fact that the oblique case begins amplifying fields much earlier in time. Similarly for the parallel benchmark simulation, even at late times, most of the upstream has $\delta B/B_0 \lesssim 1$. It is possible that this effect is due to over-injection from the downstream and subsequent unphysically high energy CR production, but it may also indicate that parallel and oblique shocks are better at converting free energy in CRs into magnetic fields at a slow but steady pace, whereas quasi-perpendicular shocks make this conversion rapidly.

By examining the way CR energy scales with time, it is possible to say whether a simulation is undergoing primarily DSA ($E \propto t$) or SDA ($E \propto t^2$). 
Figure \ref{fig:Emax_t} shows that the parallel case follows DSA predictions and scales as $t$; the quasi-perpendicular case, instead, matches SDA, growing sharply with a $t^2$ dependence. 
However, the oblique case appears to go through periods of SDA with rapid energy growth, then fall into something that looks more like DSA. 
While particles in pure SDA are typically advected after relatively few cycles of energy gain, particles undergoing a combination of fast bursts of SDA and long periods of DSA may optimize both acceleration mechanisms and grow to higher energies than would be possible from either mechanism by itself.

We also note that the duration of confinement of CRs in oblique and perpendicular shocks is not yet well understood, and it is possible that particles may drift along the shock in SDA until they have reached quasi-parallel configurations, after which they may undergo DSA (see the discussion about SN 1006 in \cite{Orusa2023}). This subtlety and a more nuanced treatment of compression ratios for oblique and quasi-perpendicular shocks will be considered in future works.

If CRs are allowed to go through periods of very rapid acceleration at oblique shocks, it may be possible to boost the expected energies well into the PeV range and fully explain the knee. 

\section{Acknowledgements}
E.S. is supported by the U.S. National Science Foundation Graduate Research Fellowship Program, grant number 2140001.
Simulations were performed on computational resources provided by the University of Chicago Research Computing Center and on  TACC's Stampede2 thourgh ACCESS (formally XSEDE) allocation (TG-AST180008).
D.C. is partially supported by NASA through grants 80NSSC20K1273 and 80NSSC18K1218 and NSF through grants AST-1909778, PHY-2010240, and AST-2009326; C.C.H was supported by NSF FDSS grant AGS-1936393 and NASA grant 80NSSC20K1273.

%
%
%

\end{document}